\documentclass{mem}
\usepackage{natbib}\usepackage{txfonts}\usepackage{balance}
\usepackage{graphicx}
\usepackage[a4paper]{hyperref}
\idline{75}{282}
\begin{document}
\def\teff{$T\rm_{eff }$}
\def\kms{$\mathrm {km s}^{-1}$}

\title{
Observational evidence for stellar mass binary black holes
and their coalescence rate
}

   \subtitle{}

\author{
T. \,Bulik\inst{1,2} 
\and K. Belczynski\inst{3,4,1}
          }

 \date{\today}

\institute{
Astronomical Observatory, The University of Warsaw, 
Aleje Ujazdowskie 4, 00-478 Warsaw, Poland
\and
Nicolaus Copernicus Astronomical Center,
Bartycka 18, 00716 Warsaw, Poland
\and
Los Alamos National Laboratory, P.O. Box 1663, MS 466, Los Alamos, NM 87545 , USA
\and 
Oppenheimer Fellow 
\\
\email{tb@astrouw.edu.pl,~kbelczyn@lanl.gov}
}

\authorrunning{Bulik }

\titlerunning{Evidence for BBHs}

\abstract{
We review the formation scenarios for binary black holes, and
 show that their coalescence rate depends very strongly on the 
 outcome of the second mass transfer. However,
the observations of IC10 X-1, an binary with a massive black
hole accreting from a Wolf-Rayet star proves that this 
mass transfer can be stable. We analyze the future
evolution of IC10 X-1 and show that it is very likely 
to form a binary black hole system merging in a few Gyrs.
We estimate the coalescence rate density of such systems 
to be $ 0.06 \, {\rm Mpc}^{-3} {\rm Myr}^{-1}$, and 
the detection rate for the current LIGO/VIRGO  of
$ 0.69 \,{\rm yr}^{-1} $, a much higher value than the 
expected double neutron star rate. Thus the first detection 
of a gravitational wave source is likely to be a coalescence 
of a binary black hole.

\keywords{Stars: binaries
 }
}
\maketitle{}

\section{Introduction}

Detecting binary black holes (BBH) is extremely difficult. 
They do not emit significant electromagnetic radiation,
like the black holes (BH) in accreting systems. They can only be discovered
via their gravitational influence: either through microlensing
or through observation of burst of gravitational waves when they 
coalesce. Microlensing caused by a BBH is an extremely rare 
event, we know only a few candidates for microlensing through single 
BHs \citep{2002ApJ...579..639B,2002MNRAS.329..349M,2005ApJ...633..914P}, and the the microlensing by BBH must be much less frequent.
The development of gravitational wave observatories on the other hand 
is very promising and may lead to detection of a BBH coalescence.
The current sensitivity of LIGO and VIRGO interferometers permits to detect 
a coalescence of a system of two $10\,$M$_\odot$ BHs at a distance exceeding
100Mpc \citep{2009PhRvD..79l2001A}. 
The sensitivity of the instruments is being improved which 
brings us closer to a possibility of detection.

In this paper we investigate discuss the possibility of existence of BBHs.
In section 2 we outline the challenge of formation of a BBH from the point 
of view of binary evolution. In section 3 we summarize the observations of IC10 X-1, and discuss its future evolution. We estimate the merger and detection rate of BBHs based on IC10 X-1 in section 4, 
and in section 5 we summarize and discuss these results.

\section{Challenges in formation of binary black holes}

The formation scenario of BBHs has been discussed in the 
literature by many authors, see e.g. \citet{1997PAZh...23..563L,1997MNRAS.288..245L,2000A&A...355..479B,2001PhyU...44R...1G,2007ApJ...664..986B}.
Here we provide an outline based 
on the StarTrack binary population synthesis code \citep{2002ApJ...572..407B,2008ApJS..174..223B}.

In order to form  a tight BBH with a merger time shorter 
than the Hubble time we need to start with binary consisting of two massive stars, 
with the mass above $20\,$M$_\odot$. They will evolve very quickly and soon one of them,
the initially more massive will enter the giant phase. This will 
lead to a mass transfer that will initially be meta stable but once the mass 
ration becomes close to unity it will stabilize. After this initial mass transfer the system will consist of a He star - the core of the initially more massive one, and the rejuvenated companion, that has gained a significant amount of mass from the companion. The Helium star evolves further and 
soon a first BH is formed. Now the system consist of a BH and the companion that 
will soon become a giant an start transferring mass onto the BH. 
This ensuing mass transfer is the bottleneck of the BBH formation. The BH
has a mass that is typically much smaller than the rejuvenated donor.
Therefore the mass transfer is not stable and the system enters a common 
envelope (CE) phase. The system may survive it yet if the donor has developed 
a well defined structure of a core and envelope \citep{2000ARA&A..38..113T}. The simulations, however, show that the donors do not have that structure and the CE phase leads to a merger
rather than formation of a tight binary. However, if a binary manages to survive this mass transfer episode than its further evolution is simple. The donor is stripped of its envelope and it becomes a He star. This star evolves also quickly and explodes forming a second BH. Thus if a system avoids 
a merger during the second mass transfer than a merging BBH can be formed. 

These considerations have been presented by \citet{2007ApJ...664..986B}. 
That paper  concludes that although BBH are very bright and detectable by gravitational wave interferometers up to very large distances the effective merger rate will be quite low since the formation rate of BBH is extremely small if the system do not survive 
the CE episode. However, our knowledge of the physics involved in 
CE evolution is poor and this leads to huge uncertainty in the calculations
of formation and merger rates od BBHs. 
This conclusion was based on the analysis of binary evolution
of stars with the solar metallicity. The evolution of metal free star 
has been considered by \citet{2004ApJ...608L..45B}.

\section{IC10 X-1}

IC10  has been discovered 120 years ago \citep{1889AN....120...33S}. 
It is an irregular dwarf galaxy at distance 
approximately 660kpc \citep{1996AJ....111..197S,1999ApJ...511..671S,2000A&A...363..130B}
It has a rather low metallicity of $0.15 Z_\odot$  \citep{1979A&A....80..155L},
and it   is the only starburst galaxy in the Local
Group, with the star formation rate of $0.04-0.08$\,M$_\odot$yr$^{-1}$  \citep{1986PASP...98....5H}.
IC10 hosts a very large number of WR stars \citep{2001A&A...370...34R,2002ApJ...580L..35M}, with 
the surface density of WR star being four times larger than 
anywhere else in the Local Group even for the regions of 
highest star formation \citep{1998ApJ...505..793M,2001A&A...366L...1R,2003A&A...404..483C}. 
The ROSAT X-ray observations \citep{1997MNRAS.291..709B}
revealed several sources, with IC10 X-1 being the brightest.

\subsection{Observations and the present state}

IC10 X-1 has already been reported to show  some X-ray variability in the 
discovery paper \citep{1997MNRAS.291..709B}.
The optical companion has been identified as a WR star MAC92
\citep{2004A&A...414L..45C}, with  the use of WR star 
catalogues obtained previously \citep{1992AJ....103.1159M,2003A&A...404..483C}.
This preliminary optical identification has been
confirmed by a detailed study with the Chandra and Hubble observations
\citep{2004ApJ...601L..67B}. Further Chandra observations 
\citep{2005MNRAS.362.1065W} have confirmed the X-ray variability 
of IC10 X-1: the X-ray flux has been monitored over a period
of half a day and it varied systematically by a factor of four.
This suggested that IC10 X-1 may be eclipsing  accretion powered  binary. 
An analysis of a long time X-ray flux using Chandra 
data together if the X-ray monitor on board of Swift satellite
revealed the orbital period of 32 hours \citep{2007ApJ...669L..21P}. 
In this paper the authors analyzed the spectroscopic 
observations of IC10 X-1 at two different epochs and obtained a lower limit 
on the radial velocity amplitude. This lead to a preliminary 
estimate of the mass of the BH in the system. The mass of the
WR star was estimated to be at least $17\,$M$_\odot$. For the inclination of 90 degrees
this implied a BH mass of $23\,$M$_\odot$.  For smaller inclinations and higher 
masses of the WR star the estimate mass of the black hole was larger.
Thus, IC10 X-1 hosts one of the most massive stellar BHs known.
A few months later a detailed spectroscopic study of this 
binary lead to measurement of a detailed radial velocity curve \citep{2008ApJ...678L..17S},
and 
the the analysis in this paper confirms 
the mass estimate of \citet{2007ApJ...669L..21P}.

Thus IC10 X-1 is a binary consisting of a BH with the mass of at least $23\,$M$_\odot$
accreting from a WR star with mass of $17\,$M$_\odot$ or more. If the mass
of the WR is larger than the BH is also more massive. For the 
upper limit on the WR star mass of $35\,$M$_\odot$ and the smallest 
possible inclination still allowing for eclipses the estimate of the mass
of the BH is $\approx 34\,$M$_\odot$.

\subsection{Future evolution}

What will be the future evolution of IC10 X-1 and what are its
implications? 
The WR donor star in the binary  is losing mass at through its strong wind. 
The mass loss rates from WR stars are highly uncertain. They have been estimated 
with various observation techniques \citep{1990A&A...231..134N,2000A&A...360..647H,2000A&A...360..227N}.
We have estimated the expected mass loss rates from  stars in low metallicity environment like the 
one in IC10 and calculated the final remnant mass as a function of the initial star mass,
for details see \citet{2009arXiv0904.2784B}.
One clue to the mass loss rates comes from the existence 
of the BH with the mass of at least $23\,$M$_\odot$. In order to form 
such a massive BH the initial star must have been very massive and 
the winds strength must not have been strong. In fact only the models
where the winds a scaled down by at least 50\% from the fiducial values 
REF lead to formation of BHs with the mass exceeding $23\,$M$_\odot$.

Assuming that the current mass loss rate is not stronger than the 
one offered for the progenitor of the BH in the system we find that the WR star will
lose a few solar masses before it collapses and forms a BH.
The mass of the BH formed in this collapse will not differ much from the 
mass of the collapsing star and will likely be $\approx 13\,$M$_\odot$, assuming 
the present mass of $17\,$M$_\odot$. If the mass of the WR star is larger
than the BH will be more massive. 

The typical lifetime of a $17\,$M$_\odot$ WR star is several hundred thousand
years, certainly not exceeding $10^6$yr. What is remarkable about this binary 
that it is a system that defies the "bottleneck" problem outlined 
in section 2. The current mass of the first formed BH is large enough
to make the mass transfer from the companion stable. Thus the apparent problem 
in formation of BBHs outlined above is automatically solved. The system 
is in a stable mass transfer state after the formation of the first black hole and its
companion is likely to form another BH within the next few hundred thousand years.

Also the binary should not be disrupted in the formation of the second BH. 
First the system is not going to loose a lot of mass when the  BH is formed.
Second, the kick velocity may be imparted on the newly formed BH. However,
numerical estimates of such kicks show that they are smaller than in the case of 
neutron stars, and may reach at most $150$km\,s$^{-1}$. 
The relative orbital velocity of the components of the binary 
has been measured to be $\approx 800$km\,s$^{-1}$. The kick velocity would have to be larger than that 
in oder to disrupt the system. Thus the system is not going to be disrupted
by the formation of the second BH.

Therefore in a few hundred thousand years IC10 X-1 will evolve into 
BBH system with the masses of at least $23\,$M$_\odot$ and 
$13\,$M$_\odot$. The orbit will most likely be similar to the current one, however it may be altered
by the natal kick. If the orbit remains the same then the merger time 
of the BBH binary will be $\approx 3 $Gyrs. A detailed simulation
with the kick velocity drawn from Maxwellian distribution with the width 
of $150$km\,s$^{-1}$, has shown that only in $3$\% of cases 
the merger time of the BBH exceeds the Hubble time.

\section{The binary black hole population}

\subsection{Estimate of the BBH merger rate density}

The estimate of the cosmic merger rate density based on a single object 
may seem like futile task yet in the field of gravitational wave astronomy 
this has already been done. The observational estimate of the BNS merger rate 
depends crucially on the observations of a single object, namely the binary pulsar
J0737-3039 \citep{2003Natur.426..531B}. 

In order to estimate the rate density we need to evaluate the volume in which 
IC10 X-1 could have been  found, as well as the time in which it is observable.
In order to estimate the volume we need to analyze critically 
the crucial observations that led us to the identification of
the nature of IC10 X-1, and find the most constraining one. 
IC10 X-1 was identified as a variable X-ray source, and X-ray observations 
allowed to determine its orbital period. The second crucial observation
was the spectroscopic measurement of the the radial velocity 
curve. This is the most constraining one: the spectroscopy 
of star is possible down to the apparent magnitude of $m_V\approx 21$.
For a WR 
with the absolute magnitude of $M\approx -5$  this corresponds to the distance modulus
of $25$, i.e. the distance of $R_{obs}=2$Mpc. Thus the volume in which 
IC10 X-1 is observable is $V_{obs}=4\pi R_{obs}^3/3\approx 33.5$\,Mpc$^3$.
In the following we will assume conservatively that the entire 
sky has been surveyed for IC10 X-1 like objects. Since IC10 X-1 has been 
discovered and identified in X-rays, the duration of the X-ray bright 
phase defined the time it is observable. For IC10 X-1 we assume conservatively
that the observability time is equal to the lifetime of 
the WR star, namely   $t_{obs}=0.5$Myr. Therefore the local 
formation rate of IC10 X-1 like objects is 
\begin{equation}
{\cal R} = (V t_{obs})^{-1}= 0.06 \, {\rm Mpc}^{-3} {\rm Myr}^{-1}\, .
\label{eq:}
\end{equation}
Since nearly all of the systems that form will merge within the Hubble time 
this is also the estimate of the merger rate density.

\subsection{The expected BBH merger rate}

The gravitational wave signal from a coalescing binary in the inspiral phase
depends on a single parameter, the chirp mass
${\cal M} = (m_1 m_2)^{3/5}(m_1+m_2)^{-1/5}$, where 
$m_i$ are the masses of the components of the binary. 
For a given signal to noise ratio the range up to 
which a binary can be detected is $D\propto 
{\cal M}^{5/6}$, and thus the volume where binaries can be detected scales 
as $V_{det}\propto {\cal M}^{5/2}$.
The chirp mass for the BBH that will form from IC10 X-1 is 
$\approx 14\,$M$_\odot$, while the chirp mass for 
the double neutron star (DNS) binary consisting of two 
$1.4\,$M$_\odot$ star is $\approx 1.2\,$M$_\odot$.
Thus the  the BBH that forms out
of IC10 X-1 from a distance   $7.8$ times larger 
than the fiducial DNS binary. For the already finished S5 run of the
LIGO detector the detectability range for DNS system was
$\approx 18$\,Mpc, so the detectability distance for a BBH with the chirp mass
of $\approx 14\,$M$_\odot$
was $ D_{BBH} \approx 140$\,Mpc. We can now estimate the expected 
BBH coalescences rate, as the volume observed was 
$V_{BBH} \approx 11.5\times 10^6$\,Mpc$^3$. Thus the expected
detection rate is
\begin{equation}
{\cal N} = 0.69 \tilde D_{DNS}^3 \tilde{\cal M}_{BBH}^{5/6} 
\tilde R_{obs}^{-3} \tilde t_{obs}^{-1}
\,{\rm yr}^{-1} 
\label{eq:ab}
\end{equation}
where $\tilde D_{DNS}=D_{DNS}/18$\,Mpc, $\tilde{\cal M}_{BBH} =
{\cal M}_{BBH}/14.3\,$M$_\odot$, $\tilde R_{obs}=R_{obs}/2\,$Mpc, and
$\tilde t_{obs}=t_{obs}/0.5\,$Myr.
One has to remember that this estimate 
 is based on a single object so in order to be fair we have 
 presented explicitly the scalings. However we will argue 
 below that this  estimate of the detection rate should be considered as
 a conservative lower limit. For a more detailed discussion
 see \citet{2008arXiv0803.3516B}.

\section{Conclusions}

The detection rate obtained above has to be treated
with some caution as it was obtained with the use of just a single object.
However, we must stress that the assumptions made in calculating it
are rather conservative. The lifetime of the WR star has been assumed to 
$0.5$\,Myr, yet for the higher masses of the star this time is even smaller,
which makes the rate density go up. Also we have been quite conservative in determining
the detectability distance for IC10 X-1 like binaries. The spectroscopy 
is difficult even for a WR star  at the distance of to IC10, but with 
the best telescopes available it might be possible up to 2 Mpc. 
If this distance is actually smaller than the value we have assumed than the
estimate of the rate density will also go up. We have also made an assumption 
that the entire sky has been surveyed for IC10 X-1 like objects. This is 
also conservative because if the actual volume surveyed was smaller 
than the rate density  has to go up again. We have also neglected the 
upward correction on the rate due to the fact that IC10 X-1 is an eclipsing
binary.

In the calculation of the detection rate we made a crucial assumption that the 
rate density estimates locally can be extrapolated to the distance of 
hundreds of Mpc. This is an assumption that is also made in case of estimates of the 
DNS coalescence detection rate. On the scales of hundreds of Mpc the 
mean galaxy density may be a little lower than in the Local Group within 
the distance of 2Mpc, yet the effect cannot be very significant. 
It may decrease  the detection rate by a factor of a  few at most.
The expected detection rate  depends strongly on the estimated chirp 
mass of the future BBH binary. This in turn depends on the 
strength of the winds and the amount of wind mass loss from the 
WR before it forms a BH. However, we have an estimate of the 
wind from the existence of the massive BH in the system, and also form observations.
These two estimates imply that the mass of the compact object 
 to be formed from the WR star 
should not be  small. It should be stressed at this point that 
the uncertainty in the mass of the BH and the ensuing uncertainty 
in the chirp mass of the BBH is the most significant source of uncertainty 
in the estimate of the BBH coalescence rate.

The astrophysical significance of this rate is quite interesting.
First, it is much higher than the double neutron star coalescence rate.
The rate is so large that the probability of detection 
in the LIGO S5 and VIRGO VSR1 data is significant.
If the data analysis of these observations does not lead to a detection
then 
the upper limits will certainly be  very intriguing.
The LIGO and VIRGO detectors are operating now 
with the enhanced sesitivity. 
Thus, it is likely that the current observational 
 runs of LIGO and VIRGO 
will finally be awarded with  a detection and the first 
source of gravitational waves will be a  coalescing BBH.

\begin{acknowledgements}
This research has been supported by the MNISW grants 1P03D00530 and
N N203~302835. TB is grateful to the hospitality of Observatoire 
de Paris in Meudon where this paper was written.
\end{acknowledgements}

\bibliographystyle{aa}


\end{document}